\documentclass[english]{appolb}
\usepackage[T1]{fontenc}
\usepackage[latin1]{inputenc}
\usepackage{array}
\usepackage[dvips]{graphicx}
\usepackage{babel}
\usepackage{multirow}

\begin{document}
\title{{\huge Justice in the Shadow of Self-Interest} {\Large }\\
{\Large An Experiment on Redistributive Behavior}%
\thanks{Presented at the 2nd Polish Seminar on Econo- and Sociophysics, Cracow, 21-22 April 2006}
}
\author{Szymon Czarnik
\address{Institute of Sociology, Jagiellonian University, ul. Grodzka 62,
31-044 Kraków \\ {\tt scisuj@o2.pl}}}
\maketitle

\begin{abstract}
By means of laboratory experiment I examine the relation between fairness
judgments made `behind the veil of ignorance' and actual behavior
in a model situation of income inequality. As the evidence shows,
when material self-interest is at stake vast majority of subjects
tends to abandon the fairness norm. Rather small regard for efficiency
is present in the data. Furthermore, as low income players go through
a sequence of games against high earners and experience changes in
income disparity, the history effect proves to override structural
characteristics of the redistribution game.

\end{abstract}

\section{Introduction}

For good or for ill, income redistribution is one of the key
functions assumed by modern governments. The underlying motive for
it, as it is prevalently professed, is either a sympathy for the
plight of the destitute, or a need to curtail the unjustifiable
inequalities though, to be sure, it is far from common agreement
what a justifiable division of income actually amounts to. To make
matters all the more harder to analyze and judge, redistribution
takes on so diverse forms that in some cases it may be nigh
impossible to determine the net effect of income transfers. To give
just one example, state-financed higher education, for a big part of
it, is a vehicle for transfering funds from lower to higher income
families \cite{0,7}. Now the aim of this paper is not to disentangle all the
intricacies of the redistributive machinery but rather to scrutinize
the link between fairness judgments about income division and actual
behavior in model situation of income disparity under controllable
conditions of laboratory experiment. In doing so, I follow the ever
widening path of research on income redistribution conducted under
the general heading of bargaining games in experimental economics
(see \cite{1,2,3,4,5}). Readers
interested in theoretical analysis of the phenomenon are referred to
\cite{6} and particularly \cite{7}.

The basis for this paper is the experiment I conducted in May 2005.
The experiment consisted of two parts, pretest and experimental proper.
All in all, 84 students participated in the first common meeting,
and of this number 72 proceeded to subsequent experimental sessions.
In total 12 six-person sessions were held and each subject participated
in one.

The aim of the first common meeting was twofold. First, all participants
were requested to fill a questionnaire on economic and political issues,
as well as they were asked to make certain decisions dealing with
division of money. Second, they were to \emph{earn} money by performing
tasks assigned to them. As I ventured on a study of \emph{redistributive
behavior} and its relation to \emph{fairness judgments}, it was crucial
that subjects' initial payoffs were earned in such a way as to make
the evaluation of fair distribution possible. I took pains to ensure
that 1) task would be reasonably easy to perform yet 2) require various
degrees of effort; 3) performance could be explicitly measured, and
4) subjects should be able to assess their effort's worth. It is especially
uncommon in vast experimental literature that condition 4 is met.
The usual way to endow human subjects with initial incomes in all
sorts of bargaining games is either by random rule, or according to
persons' scores on some kind of a test. In former case payoffs result
from sheer luck, either good or bad. In latter case one may attribute
them to subjects' performances yet it is at best unclear why particular
payoff scheme should be employed, which leaves much room for subjects'
unelicited disapproval of a scheme actually enforced by the researcher.
In my experiment gainful task consisted in decoding one, two or four
fixed-size fragments of \emph{Chronica Polonorum} by Gall the Anonym
(in its Polish translation, to be sure). Encryption method was a straightforward
injection from the character set to a set of special symbols or ideograms,
and a suitable `dictionary' was provided at the top of each sheet.
Participants' evaluation of task's worth was elicited by means of
`request for quotation' technique, i. e. all subjects were requested
to quote, independently and in secret, their prices for decoding one,
two and four sheets, and the lowest quoters were to get extra jobs
for prices they named. As a matter of fact, 92\% of the `full-time'
subjects quoted prices below the ones actually implemented and thus
we may conclude that, save 6 exceptions out of 72, subjects had no
intrinsic reason to be dissatisfied with initial payoffs they earned.

\section{Decisions behind the veil of ignorance}

Before each experimental subject actually learned whether he was to
decode one, two or four sheets, as well as before he was told that
a recompense for decoding each single sheet was 15 PLN, he had been
requested to do the following:

\begin{quote}
{\footnotesize Consider three persons, of whom Person A solved one
sheet, Person B solved two sheets, and Person C solved four sheets.
Suppose you have a total of 105~PLN to distribute among these three
persons. How much would you pay to A, B and C respectively? }{\footnotesize \par}
\end{quote}
As subjects were yet unaware of their own income positions, their
judgments were not marred by self-interested considerations and we
may regard them as indicative of their true justice preferences. Since
actual payoffs were fixed at a piecework basis it was important whether
subjective fairness norms also posited a proportional 15-30-60 PLN
division, and if not, what kind of bias they showed. In order to grasp
it, I represented payoff scale as a weightless second-class lever
of length 4 held in equilibrium by vertical force applied at the end
of it. Amounts of work (1, 2, 4) were rendered as points on the lever
arm at distances 1, 2, and 4 from the fulcrum respectively, and corresponding
`fair' payoffs were represented as loads attached to the arm at those
points. The proportionality ($\pi$) of a given fair payoff vector
was then defined as the ratio between the moments of force around
the fulcrum produced by `fair loading' and `proportional loading'
(the latter quantity being $1\times15+2\times30+4\times60=315$ units).
To be sure, proportional fairness produces $\pi$=1 whereas every
transfer from lower to higher income person increases value of $\pi$
and vice versa. As the results show, piecework solution ($\pi=1$)
is a single self-evident fairness rule, supported by 42 out of 72
`full-time' subjects (58\%)%
\footnote{Of course other payoff vectors, e.g. 17-27-61 PLN, may yield $\pi=1$
as well but it was not the case with our subjects. %
}. Other choices go both ways reaching minimum at .92 (for 20-35-50
division) and maximum at 1.11 (for 10-20-75). We dare say that actual
support for proportionality could be even higher, had the total disposable
lump sum been set at eg. 70 PLN thus making proportional division
(10-20-40) somewhat more conspicuous for non-mathematically oriented
subjects.

After making their choices on 105 PLN division, subjects were told
what was the actual price for decoding each single sheet and they
were asked to open envelopes (handed out to them at the very beginning)
in which they were to find either one, or two, or four sheets of text
to be decrypted on the spot. It was also made clear to the participants
that all payments (15 PLN show-up fee included) should be realized
only at the end of each of the forthcoming experimental sessions.

\section{The redistribution game}

\subsection{Stage game}

Redistribution game involves two players with unequal initial incomes%
\footnote{Players' initial incomes could also be equal but then there would
be no incentive for any redistribution whatsoever. The game could
also be easily transformed to include more than two players, though
in our experiment we employed only its two-person version. For a detailed
analysis of infinitely repeated two-person redistribution game see
my article \cite{8}.%
}. For convenience sake from now on I will refer to the players as
H and L (H for high and L for low income). Game consists of two consecutive
parts. First, both members of the `dyad society' are to democratically
establish the rate of redistributive income tax which benefits L at
the expense of H. Effective tax rate is an average of the rates simultaneously
and independently proposed by both players and may be anything from
0 to 100\%. However, fiscal transfer entails executive cost defined
as a percentage of total tax revenues and thus fiscal redistribution
always leads to efficiency loss. To present redistributive tax mechanism
formally, let us denote players' initial payoffs as $p_{H}$ and $p_{L}$,
their tax votes as $t_{H}$ and $t_{L}$, and the fiscal cost as \emph{C}.
Players' after-tax payoffs are then given by the following formula:

\[
p'_{i}(T)=(1-T)p_{i}+T(1-C)\bar{p}\]
where $T=\frac{1}{2}(t_{H}+t_{L})$ and $\bar{p}=\frac{1}{2}(p_{H}+p_{L})$.
As one may easily calculate, H suffers from any positive taxation,
while L benefits iff his initial payoff is small enough, viz. $p_{L}<(1-C)\bar{p}$.
In my experiment three types of `dyad societies' were formed, depending
on subjects' earnings, namely: 30-15 PLN, 60-30 PLN, and 60-15 PLN,
and there were two possible levels of executive cost: 10 or 30\%.
Under any of these conditions, player L benefits from the operation
of tax mechanism. Obviously, as far as strict material interest is
concerned, optimal decision is for L to vote 100\% and for H to vote
0\% which results in 50\% tax rate. Thus tax system may also be perceived
as a tool with which L may take over some part of H's initial income,
while at the same time part of H's income is lost due to executive
cost. L's gain and H's harm from 50\% tax redistribution for each
combination of the type of dyad and the cost level actualized in the
experiment are presented in Table 1. The efficiency of fiscal transfer,
or a portion of money taken from H by means of tax system which finds
its way to L, is given by $\tau_{L}$/$\tau_{H}$.

\begin{table}
\begin{tabular}{|>{\raggedleft}m{5.5cm}|c|c|c|c|c|c|}
\hline \multicolumn{1}{|r|}{\textbf{\small Type of dyad}} &
\multicolumn{2}{c|}{\textbf{\small 30-15 PLN}}&
\multicolumn{2}{c|}{\textbf{\small 60-30 PLN}}&
\multicolumn{2}{c|}{\textbf{\small 60-15 PLN}}\tabularnewline \hline
\multicolumn{1}{|r|}{\textbf{\small Fiscal Cost}}&\textbf{\small
10\%}& \textbf{\small 30\%}& \textbf{\small 10\%}& \textbf{\small
30\%}& \textbf{\small 10\%}& \textbf{\small 30\%}\tabularnewline
\hline\multicolumn{1}{|l|}{\small L's gain at 50\% tax
{[}$\tau_{L}${]}}&2.63&\multicolumn{1}{r|}{.38}&5.25&\multicolumn{1}{r|}{.75}&
\multicolumn{1}{r|}{9.38}&\multicolumn{1}{r|}{5.63}\tabularnewline
\hline \multicolumn{1}{|l|}{H's harm at 50\% tax {[}{\small
$\tau_{H}${]}}}& 4.88& 7.13& 9.75& 14.25& 13.13&
16.88\tabularnewline \hline \multicolumn{1}{|l|}{Efficiency of tax
transfer}& 54\%& 5\%& 54\%& 5\%& 71\%& 33\%\tabularnewline \hline
\end{tabular}
\caption{Structural characteristics of the experimental games}
\end{table}

After fiscal transfers have been realized, players have an option
to pass some portion of their after-tax income to each other (again,
decisions are taken simultaneously with no communication between players).
These voluntary transfers are costless as there is no need to employ
resources to force anyone to do what one is willing to do on one's
own. Now the final payoffs from the game are:

\[
p''_{i}(T)=p'_{i}-g_{i}+g_{j}\]
where $g_{i}$ is the lump sum given by player \emph{i} to player
\emph{j}.

Let $\tau_{L}=p'_{L}(\frac{1}{2})-p_{L}$ be the amount that L is
capable of taking over from H by voting $t_{L}=1$ and thus boosting
tax rate to $\frac{1}{2}$, and let $\tau_{H}=p_{H}-p'_{H}(\frac{1}{2})$
be the amount lost by H as a consequence. To see dilemmatic nature
of the game, suppose H considers a free transfer to L, $g_{H}=\gamma_{H}$,
such that $\tau_{L}<\gamma_{H}<\tau_{H}$ while L chooses between
tax vote of 0 and 1.

\noindent
\begin{table}
\begin{tabular}{|p{2cm}|p{2cm}|c|c|}
\hline \multicolumn{2}{|p{4cm}|}{\multirow{2}{4cm}{Values in cells
are
final payoffs $p''_{L}$, $p''_{H}$}} & \multicolumn{2}{|c|} {$\mathbf{g_{H}}$} \\
\cline{3-4} \multicolumn{2}{|c|}{} &
\textbf{\footnotesize$\mathbf{\gamma_{H}}$} & \textbf{\footnotesize
0} \\ \hline \multicolumn{1}{|c|}{\multirow{2}{*}{{\footnotesize
$\mathbf{t_{L}}$}}} & \multicolumn{1}{c|}{\textbf{\footnotesize
0}}&{\footnotesize $p_{L}+\gamma_{H}$,
$p_{H}-\gamma_{H}$}&{\footnotesize $p_{L}$, $p_{H}$} \\ \cline{2-4}
&\multicolumn{1}{c|}{\textbf{\footnotesize 1}}&{\footnotesize
$p_{L}+\tau_{L}+\gamma_{H}$, $p_{H}-\tau_{H}-\gamma_{H}$}&
{\footnotesize $p_{L}+\tau_{L}$, $p_{H}-\tau_{H}$} \\ \hline

\end{tabular}
\caption{Normal form of dychotomized redistribution game}
\end{table}

As Table 2 reveals, both players have dominant strategies: for L it
is to vote $t_{L}=1$, while for H it is to donate nothing,
$g_{H}=0$. Both players, however, should be better off if L voted
$t_{L}=0$ and H donated $g_{H}=\gamma_{H}$. This feature of the
redistribution game makes it a semblance of asymmetric continuous
prisoner's dilemma.

\subsection{Experimental procedure}

All in all, 12 experimental sessions were held, each of them including
six subjects: two low earners of 15 PLN (S1), two medium earners of
30 PLN (S2), and two high earners of 60 PLN (S4)%
\footnote{Total size of a single group had to be limited for technical reasons. %
}. After having been acquainted with the workings of redistributive
tax mechanism, subjects underwent a four-round trial game, and then
proceeded to four `real' twelve-round games (single round being the
stage game described in the previous paragraph). Subjects' initial
incomes at the beginning of each round were equal to what they had
earned during the pre-experimental meeting. To be sure, S1s in all
their games occupied L position, S4s were in H position all the time,
while S2s were in H position when playing against S1, and in L position
when playing against S4.

All games were played in local computer network and subjects could
not communicate with each other, except they were being informed on
a current basis about decisions made by the other player%
\footnote{Program was waiting until both subjects made their decisions and only
then their choices were revealed to each other. %
}. All the time they also had in view gradually updated history panel
with data on their own as well as their partner's tax votes, after-tax
incomes, free transfers and resulting final incomes for each completed
round. It was made explicitly known to the subjects that their ultimate
payoff from the experiment (in cash) would be an average over all
rounds' final payoffs so each decision they were to make would bear
financial consequences both for them and their partners.

Each subject played four 12-round redistribution games, going through
all four partner-cost configurations. Depending on order of partners
and order of fiscal costs, there were four possible histories for
each type of player. By way of example, low income player's history
of games could be one of the following (fiscal cost in parentheses):

\begin{enumerate}
\item {[}S1-\textbf{S4} (10\%) $\rightarrow$ S1-\textbf{S4} (30\%){]} $\rightarrow$
{[}S1-\textbf{S2} (10\%) $\rightarrow$ S1-\textbf{S2} (30\%){]}
\item {[}S1-\textbf{S4} (30\%) $\rightarrow$ S1-\textbf{S4} (10\%){]} $\rightarrow$
{[}S1-\textbf{S2} (30\%) $\rightarrow$ S1-\textbf{S2} (10\%){]}
\item {[}S1-\textbf{S2} (10\%) $\rightarrow$ S1-\textbf{S2} (30\%){]} $\rightarrow$
{[}S1-\textbf{S4} (10\%) $\rightarrow$ S1-\textbf{S4} (30\%){]}
\item {[}S1-\textbf{S2} (30\%) $\rightarrow$ S1-\textbf{S2} (10\%){]} $\rightarrow$
{[}S1-\textbf{S4} (30\%) $\rightarrow$ S1-\textbf{S4} (10\%){]}
\end{enumerate}
Thus subject's partners, according to their income levels, were ordered
either decendingly (histories 1 or 2) or ascendingly (histories 3
or 4). The same is the case with the cost order within the doubleheaders%
\footnote{There was no change of partners within the doubleheader, though subjects
were not informed that they played two consecutive games against the
same person. In the listing above, the-same-partner games are rendered
by means of square brackets. %
}: subjects with history 1 or 3 experienced growing cost, while subjects
with history 2 or 4 experienced diminishing cost. Important feature
of this scheme is that S1-S4 (as well as S1-S2) games are played under
two much different circumstances. In history conditions 1 and 2, we
have inexperienced S1 matched with equally inexperienced S4. However,
in history conditions 3 and 4, when it comes to S1-S4 doubleheader,
S1 has already had experience of playing against S2 in first two games---and
the same holds true for S4 who has already been playing against S2
in his first two games. In this manner, we may examine the impact
of history on redistributive behavior. Mutatis mutandis, all things
said above apply to medium and high earners as well.

\section{Results}

\subsection{Justice considerations}

First thing to be noticed in the empirical data on redistribution
game is a vast disregard for efficiency manifest in mean tax votes
made by lower earners of the dyads, which is in stark contradiction
with the data one typically gets in survey studies. Basically,
respondents claim that no transfer should be made if the efficiency
loss exceeds 20-30\% \cite{5}. Now if you recall Table 1, the most
efficient game in our set was 60-15 PLN under 10\% fiscal cost, and
in this very case efficiency loss amounted to no less than 29\%. In
all other cases transfers were even less efficient, with extreme at
60-30 PLN and 30-15 PLN under 30\% fiscal cost, in which outrageous
95\% of the transfer was bound to be lost. Now subjects' actual
behavior in redistribution game amply demonstrates the difference
between `cheap talk' and decisions made when one's monetary interest
is at stake. Table~3 shows mean tax votes, final incomes and free
transfers from H to L (corrected for the amount L sent back)
obtained in the first two games, viz. before the change of partners
took place.
\begin{table}
\noindent \begin{tabular}{|r|c|c|c|c|c|c|}
\hline
\textbf{\small Type of dyad}&
\multicolumn{2}{c|}{\textbf{\small 30-15 PLN}}&
\multicolumn{2}{c|}{\textbf{\small 60-30 PLN}}&
\multicolumn{2}{c|}{\textbf{\small 60-15 PLN}}\tabularnewline
\hline
\textbf{\small Fiscal Cost}&
\textbf{\small 10\%}&
\textbf{\small 30\%}&
\textbf{\small 10\%}&
\textbf{\small 30\%}&
\textbf{\small 10\%}&
\textbf{\small 30\%}\tabularnewline
\hline
\multicolumn{1}{|l|}{{\small Higher earner's mean tax vote (\%)}}&
\multicolumn{1}{r|}{{\small 1.33}}&
\multicolumn{1}{r|}{{\small 2.22}}&
\multicolumn{1}{r|}{{\small 1.77}}&
\multicolumn{1}{r|}{{\small 1.79}}&
\multicolumn{1}{r|}{{\small 4.43}}&
\multicolumn{1}{r|}{{\small 5.12}}\tabularnewline
\hline
\multicolumn{1}{|l|}{{\small Lower earner's mean tax vote (\%)}}&
{\small 56.73}&
{\small 40.92}&
{\small 58.82}&
{\small 53.99}&
{\small 78.55}&
{\small 74.79}\tabularnewline
\hline
\multicolumn{1}{|l|}{{\small Higher earner's final income (PLN)}}&
{\small 26.80}&
{\small 26.29}&
{\small 51.91}&
{\small 49.72}&
{\small 48.76}&
{\small 46.21}\tabularnewline
\hline
\multicolumn{1}{|l|}{{\small Lower earner's final income (PLN)}}&
{\small 16.90}&
{\small 15.80}&
{\small 35.36}&
{\small 32.76}&
{\small 23.14}&
{\small 19.80}\tabularnewline
\hline
\multicolumn{1}{|l|}{{\small Mean free transfer from H to L (PLN)}}&
\multicolumn{1}{r|}{{\small .38}}&
\multicolumn{1}{r|}{{\small .64}}&
\multicolumn{1}{r|}{{\small 2.18}}&
\multicolumn{1}{r|}{{\small 2.34}}&
\multicolumn{1}{r|}{{\small .35}}&
\multicolumn{1}{r|}{{\small .30}}\tabularnewline
\hline
\multicolumn{1}{|l|}{{\small Number of games}}&
{\small 12}&
{\small 11{*}}&
{\small 12}&
{\small 12}&
{\small 11{*}}&
{\small 12}\tabularnewline
\hline
\end{tabular}
\caption{Summary of the results from first-partner games}
{\footnotesize {*} Data from one game was lost due to technical
problems.}{\footnotesize \par}
\end{table}

Of the three types of dyads, only in 30-15 PLN condition there was
a significant impact of fiscal cost on L's voting behavior---under
heavier cost his tax claim was lesser by 16 percentage points (Wilcoxon's
matched-pairs signed-ranks test, p=0.042). Still, even here under
the extremely inefficient 30\% fiscal cost, meaning that only 1 grosz
finds its way to the low-income voter out of every 20 taken from the
high earner, L's average tax vote exceeded 40\%. It is also to be
noted that where tax transfers are extremely inefficient (30-15 and
60-30 PLN dyads under 30\% cost) one may expect free transfers from
H to contribute more to L's welfare than tax redistribution does.
Indeed, in the 30-15 dyads low earners ultimately increased their
payoff by 0.80 PLN (from initial 15.00 to 15.80) but lion's share
of that increase, 0.64 PLN, was due to free transfers on part of the
`rich'. And it is quite similar with 60-30 dyads in which 2.34 out
of 2.76 PLN increase came from H, leaving only 0.42 PLN for workings
of the tax system.

Certainly, it is also a clearly evident trait in the data that larger
relative income disparity makes the appetite for redistribution harder
to resist. In 60-15 PLN dyad, mean tax vote by L was at least 20 percentage
points higher than in the other two types of dyads. This is true both
for 10 and 30\% fiscal cost, though only under 30\% cost the difference
turns out to be statistically significant (Kruskal-Wallis test, p=0.03).

However inefficient subjects' choices might have been, they might
have still reflected their a priori distributive justice preferences.
As far as L's choices are concerned, they might have affected H's
payoffs in two ways: diminish them by positive tax vote and/or increase
them by means of a free transfer. What we are going to do now is calculate
what H's final payoff would be if it depended solely upon L's decisions
(i.e. if H had not reduced his own payoff either by a non-zero tax
vote, or by a non-zero free transfer). Then we should compare the
resulting sum with L's beliefs, elicited behind the veil of ignorance,
on what his opponent's fair payoff should be.

Suppose there was enough money to defray `fair' payoffs for both of
them while H's `fair' payoff did not exceed his actual initial income.
The former condition guarantees that in order to abide H's `fair'
payoff, L did not have to give up a bit of his own `fair' payoff,
while the latter condition ensures that following L's fairness rule
demanded of him no more than preserving status quo. As a matter of
fact, was H's `fair' payoff smaller than his initial income in the
game, fairness rule would even allow a certain dose of tax redistribution.
Under such circumstances, if L's actions had driven his partner's
payoff below the 'fair' level, then it would clearly indicate that
some other considerations, be it self-interest or envy, have been
at work here that ultimately turned out to override any regard for
fairness. Now in 29 out of total 35 games that were played as first
in a series of four, the sum of initial incomes was actually greater
or equal to the sum of L's and H's `fair' payoffs as perceived by
L, and in none of these games H's `fair' payoff exceeded his initial
income. As results show, in 23 out of those 29 games subjects in L
position (79\%) violated their own fairness rules. In 60-15 PLN dyads,
which offered strongest material incentives for L to exploit tax system,
majority of low earners went as far as diminishing H's `fair' payoff
by more than 10 PLN.

\subsection{History effect}

In previous paragraph we analyzed only first-partner games. Let us
now examine how first-partner experience affected voting behavior
of 15 PLN earners in their second-partner games. When we consider
all 15-60 games, half of them was played with no experience at all,
and half of them was played after both L and H had already played
a doubleheader against 30 PLN earners. Similarly, half of all 15-30
games were played by inexperienced subjects, while the other half
was played by subjects who had already been through a doubleheader
vs. 60 PLN earners. Now, as the evidence shows, impact of history
is virtually overwhelming.

\begin{figure}[!h]
\begin{center}
\includegraphics[width = 0.99\textwidth]{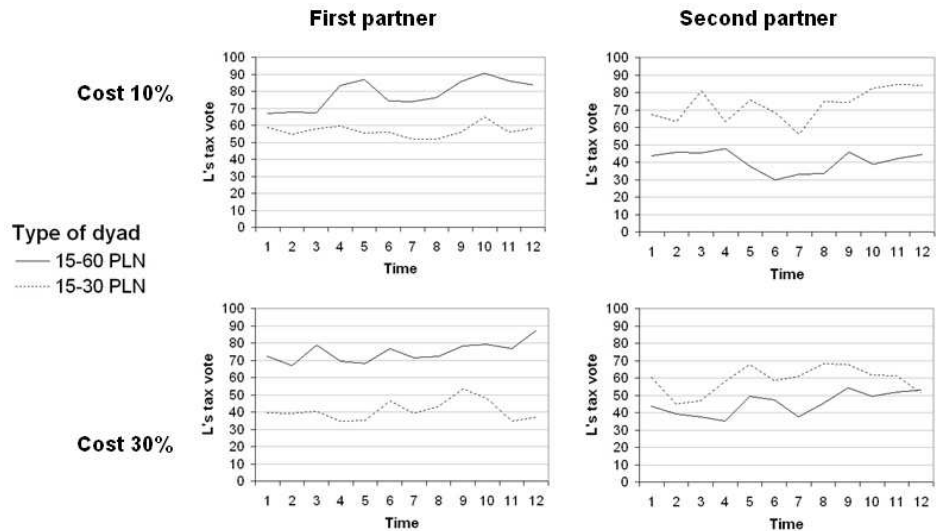}
\caption{Impact of history on low earner's tax voting}
\end{center}
\end{figure}
\vspace{-0.5cm}

In first-partner games, as it was mentioned before, larger income
disparity resulted in low earners's greater demand for tax
redistribution, and it was especially visible under 30\% fiscal
cost. Now the results from the second-partner games are in striking
contrast to this: appetite for redistribution grows substantially in
15-30 games and it lessens in 15-60 games so that eventually the
relation between income disparity and demand for fiscal
redistribution is reversed. A plausible explanation of the
phenomenon is that in terms of inertia present in social systems. In
spite of a significant change in environment, subjects tended to
inherit, to some extent, their behavior from before the change. Low
earners who played their first two games against subjects four times
richer than themselves learned that aggressive tax voting pays off
and their moral reluctance to exploit the better-off party, if any
reluctance they had indeed, must have been weakened. Thus when in
the last two games they were confronted with subjects merely twice
richer, it was relatively easier for them to forego fairness
considerations and act on their self-interest. The reverse is true
in case of low income players whose first experience was with 30 PLN
earners: it was neither so beneficial to indulge in tax
redistribution, nor it seemed so much justified to exploit subjects
whose initial income was medium rather than high. In this manner low
earners, so to say, learned to contain their appetite for
redistribution at the expense of the richer party, and it was
relatively easier for them to stick to the fairness rule once they
came up against 60 PLN Croesus. However plausible this a posteriori
explanation may seem, it is still surprising that the history effect
was so powerful as to completely override the incentives inherent in
the structure of the games.

\section{Concluding remarks}

The outright dominance of self interest over a priori justice considerations
is readily observable in the contrast between choices made behind
the veil of ignorance and those made in the course of the game. Subjects
in lower income position were taking advantage of redistributive fiscal
mechanism even in face of the gravest efficiency loss. It is a rather
sorry information for those who would like to infer people's attitudes
towards tax system from their responses to survey questions.

As far as dynamics of redistribution are concerned, we may sketch
two broad conclusions. First, substantial amount of inertia is present
in social system, and one should not expect immediate shift in behavioral
patterns even under heavy structural change. Second---as risky as
real-life generalizations made on the basis of a controlled laboratory
experiment may be---it seems that while in societies with growing
income inequalities constituencies may for some time stick to moderate
redistribution typical for earlier periods, in societies where stark
inequalities exist it may hardly be expected that income equalization
should lead to decrease in actual demand for redistribution.

The research and writing of this paper has been financed by the Polish Ministry of Education and Science as a part of the project 1~H02E~046~28.

\end{document}